# Topologically protected edge states in time photonic crystals with chiral symmetry

Yukun Yang[1,†], Hao Hu[1,†,*], Liangliang Liu[1,†], Yihao Yang[2,3,4], Youxiu Yu[1], Yang Long[5], Xuezhi Zheng[6], Yu Luo[1,*], Zhuo Li[1,*] and Francisco J. Garcia-Vidal[7]

[1]*National Key Laboratory of Microwave Photonics, College of Electronic and Information Engineering, Nanjing University of Aeronautics and Astronautics, Nanjing 211106, China*

[2]*State Key Laboratory of Extreme Photonics and Instrumentation, ZJU-Hangzhou Global Scientific and Technological Innovation Center, Zhejiang University, Hangzhou 310027, China*

[3]*International Joint Innovation Center, The Electromagnetics Academy at Zhejiang University, Zhejiang University, Haining 314400, China*

[4]*Key Lab. of Advanced Micro/Nano Electronic Devices & Smart Systems of Zhejiang, Jinhua Institute of Zhejiang University, Zhejiang University, Jinhua 321099, China*

[5]*Division of Physics and Applied Physics, School of Physical and Mathematical Sciences, Nanyang Technological University, Singapore 637371, Singapore*

[6]*Department of Electrical Engineering, KU Leuven, Leuven 3001, Belgium*

[7]*Departamento de Física Teórica de la Materia Condensada and Condensed Matter Physics Center (IFIMAC), Universidad Autónoma de Madrid, Madrid E-28049, Spain*

[†]These authors contribute equally.
[*]Corresponding authors: Hao Hu; Yu Luo; Zhuo Li
**Email:** hao.hu@nuaa.edu.cn (Hao Hu); yu.luo@nuaa.edu.cn (Yu Luo); lizhuo@nuaa.edu.cn (Zhuo Li)

**Abstract**

**Time photonic crystals are media in which their electromagnetic parameters are modulated periodically in time, showing promising applications in non-resonant lasers and particle accelerators, among others. Traditionally utilized to study space photonic crystals, topological band theory has also been translated recently to analyze time photonic crystals with time inversion symmetry, enabling the construction of the temporal version of topological edge states. However, temporal disorder can readily break time inversion symmetry in practice, hence likely destroying the edge states associated with this type of time photonic crystals. To overcome this limitation, here we propose a new class of time photonic crystals presenting chiral symmetry instead, whose edge states exhibit superior robustness over the time-reversal-symmetry-protected counterparts. Our time photonic crystal is equivalent to a temporal version of the Su-Schrieffer-Heeger model, and the chiral symmetry of this type of time photonic crystals quantizes the winding number defined in the Bloch frequency band. Remarkably, random temporal disorders do not impact the eigenfrequencies of these chiral-symmetry-protected edge states, while instead enhancing their temporal localizations. Our findings thus provide a promising paradigm to control field amplification with exceptional robustness as well as being a feasible platform to investigate various topological phases in time-varying media.**

**Table of Contents**

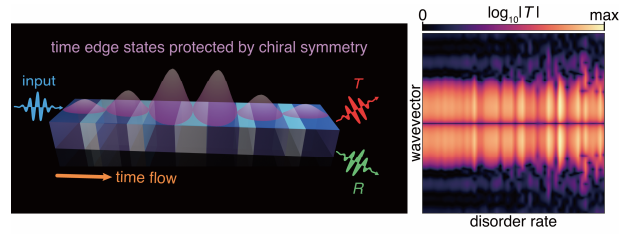

# 1. Introduction

Time-varying media refer to materials whose properties change suddenly or continuously with time. [1–6] They have attracted growing attention recently in many fields ranging from electromagnetism, photonics, acoustics and thermal physics.[7–10] Compared to their stationary counterparts, time-varying media exhibit exotic properties such as an intrinsically non-Hermiticity and non-conserved frequency,[11–13] leading to applications in magnetic-free optical isolators/circulators,[14–16] time-domain field localization[17] and time-reversal mirrors.[18] Among all time-varying media, time photonic crystals (TPCs) are of particular interest.[19–26] TPCs are materials in which the electromagnetic parameters (e.g., dielectric permittivity and/or magnetic permeability) are periodically modulated in time.[20] The periodic modulation of electromagnetic parameters results in the mutual interference between temporal reflection and refraction, leading to the formation of dispersion bands. TPCs are fundamentally different from the well-known space photonic crystals (SPCs) as the bands of TPCs are gapped in momentum (corresponding to momentum gaps), while those of SPCs are gapped in energy/frequency (corresponding to energy gaps).[22] Interestingly, momentum gaps of TPCs enable significant wave amplification.[20] This remarkable feature is entirely different from the energy gap of SPCs, in which wave transmission is forbidden inside the bandgap. Due to the ability to enhance electromagnetic waves, TPCs have already found practical applications in active metasurfaces,[19] non-resonant lasers,[20] and momentum-controlled optical amplifiers,[25] among others.

In parallel, the topological nature of crystals has been widely explored in the past decades.[27–33] Topology refers to a mathematical concept that describes a quantity that remains unchanged under continuous deformation.[34]

In the 1980s, the concept of topology was extended to physics, and this extension has deepened our understanding of material classifications[35,36] as it provided a powerful tool for analyzing and predicting material behaviors, leading to the discovery of various topological states showing remarkable robustness against external variations.[37,38] Owing to the similarities between quantum systems and classical wave systems, topological physics has been further extended to analyze electromagnetic systems such as photonic crystals.[39,40] Recently, by regularization approaches, the topology of materials even with strong dispersion could be specified.[41,42] Importantly, topological phases in photonic crystals enable a wealth of exotic phenomena including unidirectional flow,[43] valley-dependent wave transport[44], Causality-suppressed coupling[45] and spin-dependent wave transfer,[46] which allow enhancing the robustness of various photonic devices, such as waveguides,[47] lasers,[48] and transmission lines.[49]

In the past few years, the topological nature of TPCs has been explored actively. It has been theoretically revealed that the dispersion bands of *TPCs* with time inversion symmetry are associated with quantized Zak phases,[31] resulting in the emergence of topologically protected temporal edge states between two TPCs with opposite Zak phases. Lately, TPCs with the specific zak phase distribution can be readily designed using the deep learning theory.[50] However, the resonant frequencies and wave functions of these edge states under the protection of time inversion symmetry are very sensitive to temporal disorder,[51] such as perturbations in the refractive index and/or time duration and, therefore, the behaviors of these edge states can be easily affected by disorders in practice.[49] The rise time of optical responses, the fabrication imperfections and intensity oscillations of pumps would inevitably impose perturbations to TPCs that will deteriorate the temporal edge states, preventing their practical applications in information storage and transportation.

Here we propose a novel type of topological phases in TPCs that can enable temporal topological edge states with enhanced robustness. Inspired by the Su-Schrieffer-Heeger (SSH) model in real space, we design our TPCs such that the ratio between the time duration and the refractive index of each time slab within the unit cell keeps constant. As a result, our temporal version of the SSH model mimics the spinless fermions hopping with staggered hopping amplitudes in time, and exhibits chiral symmetry, as evidenced by the quantized winding number defined in the frequency Brillouin zone. The quantized winding numbers allow us to predict the emergence of temporal topological edge states, which are localized in the time domain wall between two TPCs with distinct winding numbers. Due to the existence of a temporal topological edge state, the transmission spectrum shows a sharp dip in the middle of the momentum gap where the transmission is otherwise amplified. As protected by the chiral symmetry, the resonant frequencies and associated wavefunctions of the resulting temporal edge states are highly robust against external perturbations.

## 2. Results and discussion

Without loss of generality, we consider a TPC whose refractive index is modulated periodically in time but remains homogeneous in space [see a schematic picture in Fig. 1(a)]. Our studied material should be dispersionless, because in the dispersive system, the frequency-dependent refractive index readily breaks the chiral symmetry of our system. A unit cell of this TPC comprises two temporal interfaces and three time slabs where time interfaces are formed by ultrafast variations of the refractive indices of time slabs in time.[13] We define the time slab between the two interfaces in the same unit cell as internal time slab, and that between two interfaces in neighboring unit cells as external time slab. The refractive index and time duration of the internal (external) time slab are denoted as $n_1(n_2)$ and $t_1(t_2)$, such that the period of the TPC is $t_B = t_1 + t_2$. Due to the translational symmetry in space, the momentum, denoted as $k$, is a conserved quantity. As such, the frequencies of the reflected and transmitted light are related to the incident frequency as $\omega_{r(i,3-i)} = -n_i\omega_i/n_{3-i}$ and $\omega_{t(i,3-i)} = n_i\omega_i/n_{3-i}$, respectively, where $i = 1,2$; $(i, 3-i)$ denotes the boundary between the media of refractive index $n_i$ and $n_{3-i}$. By imposing time boundary conditions, i.e., continuity of both magnetic flux density, **B**, and electric displacement, $\boldsymbol{D}$, the relation between electromagnetic fields at two neighboring time interfaces can be written as,

$$\begin{bmatrix} D_{j-1,j} \\ B_{j-1,j} \end{bmatrix} = \begin{bmatrix} \cos(kc\bar{t}_j) & i\frac{n_j}{\eta_0}\sin(kc\bar{t}_j) \\ i\frac{\eta_0}{n_j}\sin(kc\bar{t}_j) & \cos(kc\bar{t}_j) \end{bmatrix} \begin{bmatrix} D_{j,j+1} \\ B_{j,j+1} \end{bmatrix}, \tag{1}$$

where $\eta_0$ refers to the wave impedance in vacuum, $\bar{t}_j = t_j/n_j$ represents the ratio between the temporal duration and the refractive index at the $j^{th}$ slab, and $D_{j-1,j}(B_{j-1,j})$ accounts for the y-polarized electric displacement (z-polarized magnetic flux density) at the boundary between the $(j-1)^{th}$ and $j^{th}$ slabs.

To find the equation for the eigenfunctions of our system, we sort the Eq. (1) into an expression by eliminating the magnetic flux density $B$ as,

$$\frac{D_{j-2,j-1}}{\frac{n_{j-1}}{\eta_0}\sin(kc\bar{t}_{j-1})} + \frac{D_{j,j+1}}{\frac{n_j}{\eta_0}\sin(kc\bar{t}_j)} = \left( \frac{\cos(kc\bar{t}_{j-1})}{\frac{n_{j-1}}{\eta_0}\sin(kc\bar{t}_{j-1})} + \frac{\cos(kc\bar{t}_j)}{\frac{n_j}{\eta_0}\sin(kc\bar{t}_j)} \right) D_{j-1,j}, \tag{2}$$

which relates electric displacement $D$ in three neighboring boundaries. Notably, when the refractive indices and temporal durations fulfill $t_1/n_1 = t_2/n_2 = \bar{t}$, Eq. (2) can be further simplified into a group of linear equations for Bloch modes in the TPC,

$$\begin{aligned} &\vdots \\ g_{\text{inter}}D_{j-3,j-2} + g_{\text{intra}}D_{j-1,j} &= \lambda D_{j-2,j-1} \\ g_{\text{intra}}D_{j-2,j-1} + g_{\text{inter}}D_{j,j+1} &= \lambda D_{j-1,j} \\ g_{\text{inter}}D_{j-1,j} + g_{\text{intra}}D_{j+1,j+2} &= \lambda D_{j,j+1} \\ &\vdots \end{aligned} \tag{3}$$

where $g_{\text{intra}} = n_1^{-1}/(n_1^{-1} + n_2^{-1})$, $g_{\text{inter}} = n_2^{-1}/(n_1^{-1} + n_2^{-1})$ and $\lambda = \cos(kc\bar{t})$. Notably, Eq. (3) is equivalent to a matrix eigenvalue equation as follows,

$$\begin{bmatrix} \ddots & \vdots & \vdots & \vdots & \vdots & \vdots & \cdot \\ \cdots & 0 & g_{\text{inter}} & 0 & 0 & 0 & \cdots \\ \cdots & g_{\text{inter}} & 0 & g_{\text{intra}} & 0 & 0 & \cdots \\ \cdots & 0 & g_{\text{intra}} & 0 & g_{\text{inter}} & 0 & \cdots \\ \cdots & 0 & 0 & g_{\text{inter}} & 0 & g_{\text{intra}} & \cdots \\ \cdots & 0 & 0 & 0 & g_{\text{intra}} & 0 & \cdots \\ \cdot & \vdots & \vdots & \vdots & \vdots & \vdots & \ddots \end{bmatrix} \begin{bmatrix} \vdots \\ D_{j-3,j-2} \\ D_{j-2,j-1} \\ D_{j-1,j} \\ D_{j,j+1} \\ D_{j+1,j+2} \\ \vdots \end{bmatrix} = \lambda \begin{bmatrix} \vdots \\ D_{j-3,j-2} \\ D_{j-2,j-1} \\ D_{j-1,j} \\ D_{j,j+1} \\ D_{j+1,j+2} \\ \vdots \end{bmatrix}. \tag{4}$$

Note that the eigenvalues in this equation are related to the momentum $k$ by means of $\lambda$, and the eigenfunctions in our TPC refer to fields at discrete time intervals and, therefore, the magnitudes $g_{\text{intra}}(g_{\text{inter}})$ represent the intra-cell (inter-cell) coupling between sites at the nearest time moments. Now we demonstrate that, as in the original SSH model, our TPC also displays a chiral symmetry. To illustrate this, we first calculate the mode dispersion of the eigenvalues by taking the Fourier transform of the eigenfunctions in time, such that the real-space Hamiltonian is transformed into a 2×2 frequency-space Hamiltonian as,

$$H(\omega_B) = \begin{bmatrix} 0 & g_{\text{intra}} + g_{\text{inter}} e^{i\omega_B t_B} \\ g_{\text{intra}} + g_{\text{inter}} e^{-i\omega_B t_B} & 0 \end{bmatrix}, \quad (5)$$

whose eigenvalue is $\lambda_{\omega_B} = \cos(kc\bar{t})$. The Hamiltonian $H(\omega_B)$ exhibits chiral symmetry, because our system satisfies $\sigma_z H(\omega_B)^\dagger \sigma_z^{-1} = -H(\omega_B)$, where $\sigma_z$ refers to the third Pauli matrix. By solving the eigenvalue problem, one can obtain the dispersion relation of the eigenfunctions as,

$$\lambda_{\omega_B} = \pm\sqrt{g_{\text{intra}}^2 + g_{\text{inter}}^2 + 2g_{\text{intra}}g_{\text{inter}}\cos(\omega_B t_B)}. \quad (6)$$

The band structure of this discretized model for the TPC is plotted in Fig. 1(d), with the parameters being $n_1 = 4$, $n_2 = 2$, $\bar{t} = 1$ fs, and $t_B = 6$ fs. This discretized model can be regarded as a two-state system, leading to only two branches of dispersion bands. In addition to the momentum bandgap as previously revealed in TPCs,[20,22,23] these two bands are symmetric with respect to the chiral point $\lambda_{\omega_B} = 0$ [see green dashed line in Fig. 1(d)]. The existence of a chiral point is a key signature of a system with the chiral symmetry. By replacing the $\lambda_{\omega_B}$ by $k$ in Eq. (6), we obtain the realistic dispersion relation, corresponding to the band structure of the continuous model [Fig. 1(c)]. The band structure of the continuous model contains infinite dispersion bands, which are periodically arranged in the momentum space. As a result, bandgaps periodically emerge at $k/k_0 = (2m + 1)$, where $k_0 = \pi/(2c\bar{t})$ and $m$ is an integer [see light blue region in Fig. 1(c)]. Since all bandgaps in the continuous model can be transformed into the one in the discretized model, the topological phase of the bandgap in $\lambda_{\omega_B}$ applies to all the momentum gaps of the TPC with chiral symmetry [see Section S3 of the Supplemental Information].

TPCs with chiral symmetry exhibit an integer topological invariant, i.e., the winding number. Inspired by the conventional theory associated with the SSH model,[52] the winding number in TPCs can be defined in the frequency Brillouin zone as,

$$w = \frac{1}{2\pi} \int_0^{2\pi} \partial_{(\omega_B t_B)} \arctan \frac{P_y}{P_x} d(\omega_B t_B),  \qquad (7)$$

where $P_x = \langle\psi|\sigma_x|\psi\rangle$, $P_y = \langle\psi|\sigma_y|\psi\rangle$ are the two components of the pseudospin vector $\bar{P}$. Moreover, $\sigma_x$ and $\sigma_y$ are the first and second Pauli matrices and $|\psi\rangle$ refers to the eigenfunction of the frequency-space Hamiltonian matrix $H(\omega)$. Following such a methodology, we have obtained the winding number for two types of unit cells, i.e., the one with $n_1 = 4$ and $n_2 = 2$, the other with $n_1 = 2$ and $n_2 = 4$, respectively. When $n_1 > n_2$ (i.e., $g_{\text{intra}} < g_{\text{inter}}$) [see Fig. 2(a)], the endpoint of the pseudospin vector encircles the origin one time on the equatorial plane of the Bloch sphere as the Bloch frequency $\omega$ continuously travels along the Brillouin zone. In this case, the winding number is 1, indicating a nontrivial topological phase. On the other hand, when $n_1 < n_2$ (i.e., $g_{\text{intra}} > g_{\text{inter}}$) [see Fig. 2(b)], the endpoint of the pseudospin vector fails to encircle the origin. As such, the winding number is 0, showing a trivial topological phase. Notably, these winding numbers of the TPC and its spatial counterpart are different, even if the unit cell has an identical refractive index distribution. This is due to the space-time duality, as the time photonic crystals with refractive index $n(t)$ correspond to the space photonic crystals with refractive index $1/n(x)$ (see more details in Section S2 of the Supplemental Information).

The presence of bands with different winding numbers allows for the design of TPC-based structures that could support temporal topological edge states. To demonstrate this, we construct a system by interfacing two TPCs with chiral symmetry [see Fig. 3(a)]. For convenience, we will denote the TPC at the bottom of the panel as TPC1, and the one at the top as TPC2. Due to different parameter setups, the winding number of TPC1 is always 1, while that of TPC2 depends on $n_x$. By changing $n_x$ from 1 to 9, the intra-coupling in TPC2 is reduced, resulting in a topological phase transition occurring at $n_x = n_1 = 4$, where the winding number changes from $w = 0$ to $w = 1$ [see Fig. 3(b)]. Therefore, when $n_x > 4$, the winding numbers of TPC1 and TPC2 are identical (i.e., $w = 1$) and no temporal edge state is supported. In this regime, the transmission spectrum of the TPC-based structure shows a prominent field enhancement in the momentum bandgap [see Fig. 3(c) and black line in Fig. 3(d)], as the growing mode located inside the momentum bandgap is always the dominant one [see black line in Fig. 3(e)]. On the other hand, when $n_x < 4$, the winding numbers of the two TPCs are distinct, and a temporal topological edge state emerges at the chiral point [see green dashed line in Fig. 3(c)]. The emergence of this edge state results in a sharp dip in the transmission spectrum located inside the momentum bandgap [see red line in Fig. 3(d)]. More specifically, the emergence of temporal topological edge states enforces the attenuating gap mode to be the dominant one after the time domain wall [see red line in Fig. 3(e)], and thus, dramatically reduces the transmission at the eigenvalue of the temporal topological edge state.

Finally, we demonstrate that the temporal topological edge states emerging in this type of chiral TPCs are distinct from those supported by TPCs that present time inversion symmetry [The time inversion symmetry refers to the one that inversing the unit cell of a time photonic crystal in time, the unit cell remains unchanged, i.e., $n(t) = n(-t)$].[31] To reveal the difference, we investigate the influence of disorders on the eigenvalues of temporal topological edge states protected by the chiral symmetry and time inversion symmetry. We introduce disorders in TPCs by randomly varying the $n_j$ of each time slab within the range of $[(1-\rho)n_j, (1+\rho)n_j]$, where the disorder rate $\rho$ is defined as the maximum variation strength of refractive indices $n_j$ in each slab of TPCs over the initial value. To make a fair comparison, we fix the value of $\bar{t} = t_j/n_j$ by adjusting the corresponding time duration $t_j$ for TPCs with chiral symmetry, while the relation of $n(t) = n(-t)$ is kept unchanged for TPCs with time inversion symmetry when applying disorders. As shown in the spectrum in Fig. 4(a), the eigenvalue of chiral-symmetry-protected temporal edge states is very robust to disorders even when $\rho$ is up to 0.5 [Fig. 4(a&c)]. In sharp contrast, the eigenvalue of temporal edge states protected by time inversion symmetry oscillates drastically as $\rho$ increases from 0 to 0.5 [Fig. 4(b&d)]. Moreover, to show the impact of disorders on the field distribution of the two distinct types of temporal topological edge states, we choose 5 typical values of $\rho$ and depict the corresponding electric displacement distribution in Fig. 4(e, f). In the system with chiral symmetry, the temporal topological edge states are always strongly localized at the temporal interface between the two TPCs. Interestingly, the localization is even stronger as $\rho$ increases towards 0.5. However, in the structure with time inversion symmetry, the increase of disorder rate significantly weakens the localization of the temporal topological edge states. This difference indicates that the disorder, which is usually unwanted, could be a merit to enhance the chiral-symmetry-protected edge states in TPCs (the statistical results are presented in supplemental information S5). Moreover, we consider a more realistic case without fixing $t_j/n_j$ to be a constant value (seeing details in the supplemental information S4), we find that the behaviors of topological edge states are still much more robust than those in ordinary time photonic crystals.

## 3. Conclusion

In conclusion, we have constructed TPCs that present a chiral symmetry. By generalizing the concept of winding number, we reveal a novel topological phase in this type of TPCs, which allows the emergence of chiral-symmetry protected temporal edge states. As a key advantage over edge states based on TPCs with time inversion symmetry, these new temporal topological edge states are much more robust against disorders. We highlight that our proposed system could be experimentally realized on various platforms, such as transmission lines[13] (seeing detailed experimental scheme in supplemental information S7) and fiber loop systems.[53] Our work not only could find practical applications in e.g., tunable optical amplifier with topologically enabled robustness, but also inspire future exploration of topological physics in high-dimensional spatiotemporal photonic crystals where the material parameters are periodically modulated in both space and time.[54]

**Acknowledgements**

We appreciate the valuable insights provided by Prof. Jiang Xiong at University of Electronic Science and Technology of China. H.H., L.L., Y.L., and Z.L. acknowledge financial support by National Natural Science Foundation of China (Grants No. 12404363, 61871215, 61771238, 61701246), Natural Science Foundation of Jiangsu Province (Grants No. BK20241374) and Distinguished Professor Fund of Jiangsu Province (Grants No. 1004-YQR23064, 1004-YQR24010), the National Key Research and Development Program of China (Grant No. 2022YFA1404903) and Fundamental Research Funds for the Central Universities, NUAA (Grant No. NS2023022, NS2024022); F.J.G.-V. acknowledges financial support by the Spanish Ministry for Science and Innovation-Agencia Estatal de Investigacion (AEI) through [Grants No. PID2021-125894NB-I00 and CEX2018-000805-M (through the Maria de Maeztu program for Units of Excellence in R&D)].

**Supporting Information**

Supporting Information Available: < Details about the Derivation of Su-Schrieffer-Heeger (SSH) model, distinction of topological phases, the topological transitions of time photonic crystals with chiral symmetry, the influence of disorders on temporal topological edge states, field localization at time edge states and the supplement on practical considerations>

This material is available free of charge via the Internet at http://pubs.acs.org

**Notes**

The authors declare no competing financial interest.

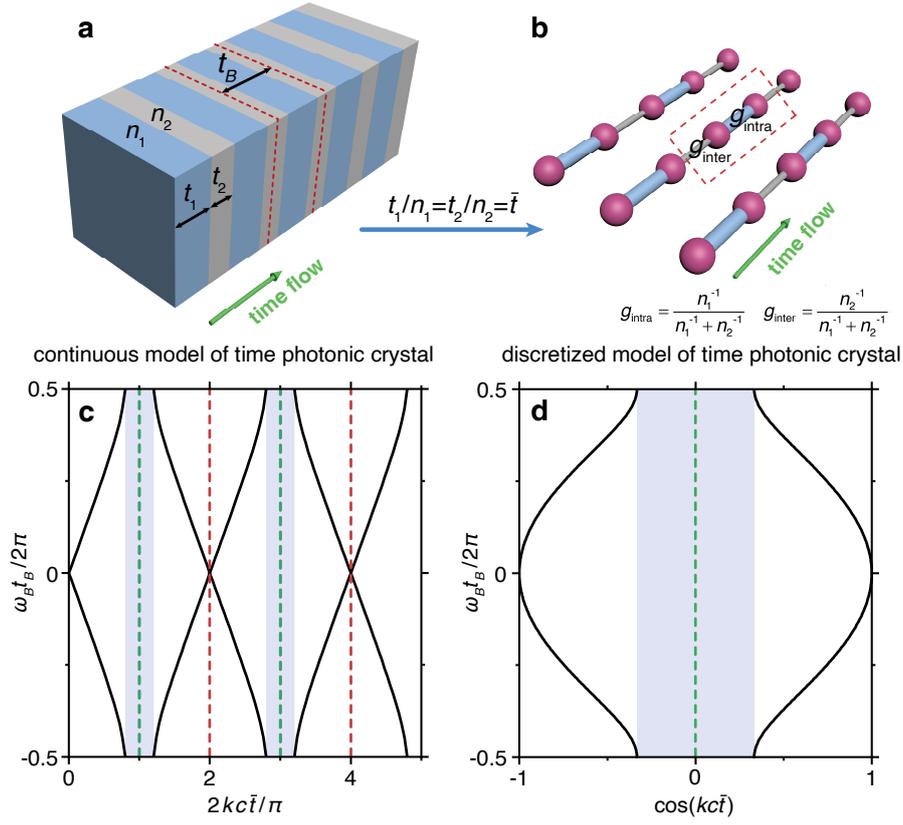

**Figure 1.** Schematic of the time photonic crystal with chiral symmetry. (a) Continuous model of TPCs. As highlighted in red-dashed lines, a unit cell comprises two temporal boundaries. The refractive index and time duration of the internal (external) time slab are $n_1(n_2)$ and $t_1(t_2)$, respectively. The period of the TPC is thus $t_B = t_1 + t_2$. (b) Discretized model of a TPC. Under the condition of $t_1/n_1 = t_2/n_2 = \bar{t}$, the continuous model of the TPC is equivalent to a discretized model preserving chiral symmetry. In such a model, $g_{\text{intra}} = 1/3$ and $g_{\text{inter}} = 2/3$ correspond to the intra-coupling and inter-coupling strengths, respectively, between the nearest sites arranged in the time-axis. (c) Dispersion relation between $\omega$ and $k$ for the TPC with chiral symmetry as obtained with the continuous model. (d) Dispersion relation between $\omega$ and $\cos(kc\bar{t})$ for the discretized model of the TPC with chiral symmetry. In (c, d), the momentum gap is highlighted in blue while the wavenumber of the Dirac point (chiral point) is indicated by the red-dashed (green-dashed) line.

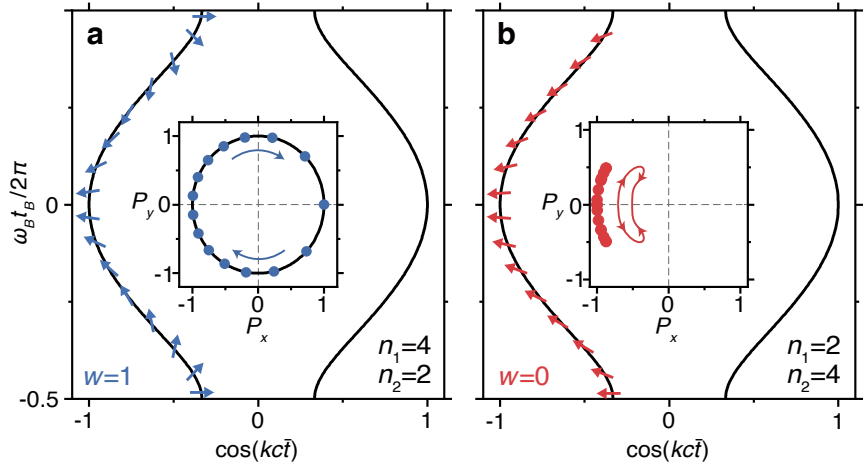

**Figure 2.** Winding number of the time photonic crystal with chiral symmetry. (a, b) Dispersion relation and winding number of the TPC with the chiral symmetry. In (a), $n_1 > n_2$, and in (b), $n_1 < n_2$. In all the panels, the insets plot the paths of the endpoints of the pseudospin vector $\bar{P} = P_x \hat{x} + P_y \hat{y}$ on the $P_x O P_y$ plane for the lowest band, where $P_x = \langle \psi | \sigma_x | \psi \rangle$ and $P_y = \langle \psi | \sigma_y | \psi \rangle$ are the two components of the pseudospin vector $\bar{P}$; $\sigma_x$ and $\sigma_y$ are the first and second Pauli matrices; and $|\psi\rangle$ refers to the periodic function of Bloch modes. The calculated pseudospin vectors are marked in all the lowest dispersion bands, with the corresponding vector components plotted in the inset. The calculated winding number is presented at the bottom-left corner in each panel.

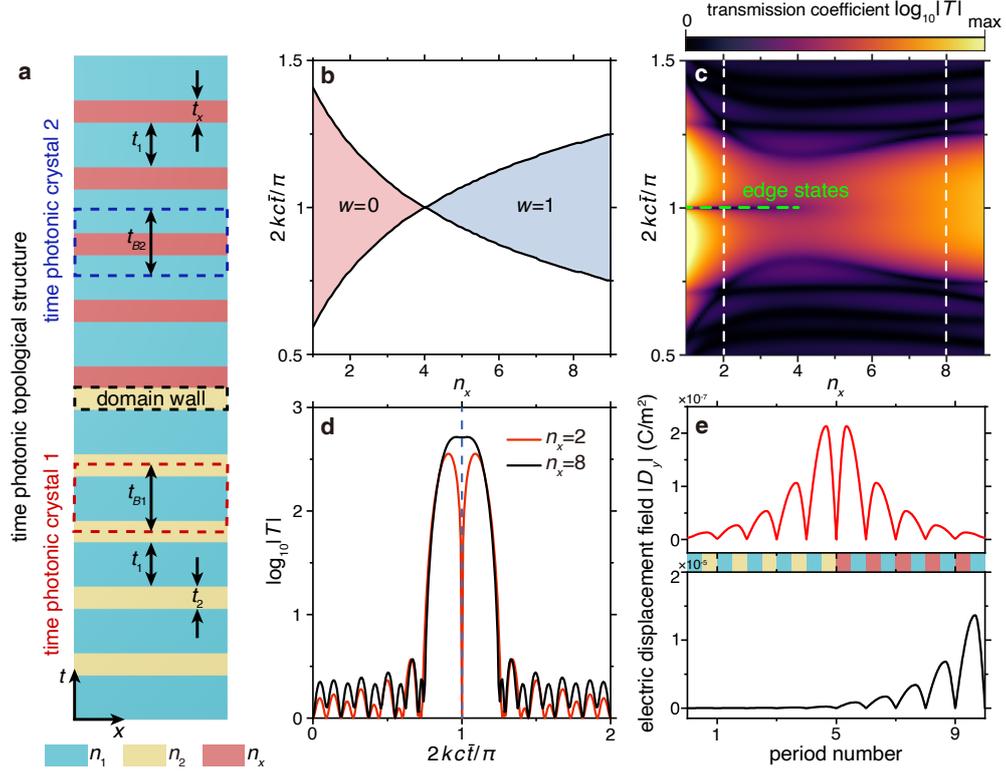

**Figure 3.** Topological transition and bulk-edge correspondence in the time photonic crystals with chiral symmetry. (a) TPC-based structure constructed by two TPCs with chiral symmetry. The unit cell in TPC1 is highlighted in a red-dashed box. The refractive index and time duration of internal (external) time slab are $n_1(n_2)$ and $t_1(t_2)$, respectively. Here, we set $n_1 = 4$ and $n_2 = 2$, and the unit cell in TPC2 is highlighted in a blue-dashed box. The refractive index and time duration of the internal (external) time slab are $n_x(n_1)$ and $t_x(t_1)$, respectively. (b) Topological phase transition of TPC2. When $n_x < 4(n_x > 4)$, the corresponding winding number is 0(1), implying a topological transition occurring at $n_x = 4$. (c) Transmission spectrum of the TPC-based structure as a function of $n_x$ and momentum. The eigenvalue of the temporal edge state is highlighted with a green dashed line. (d) Comparison of transmission spectra of the time photonic crystals for the cases $n_x = 2$ (red line) and $n_x = 8$ (black line). (e) Comparison of electric displacement distributions, for the cases $n_x = 2$ (red line) and $n_x = 8$ (black line). The incident momentum is set as $2kc\bar{t}/\pi = 1$. Moreover, the refractive index distributions in the normalized time domain $\bar{t}$ are highlighted with different colors [see the structure between the two subpanels of Figure 3(e)].

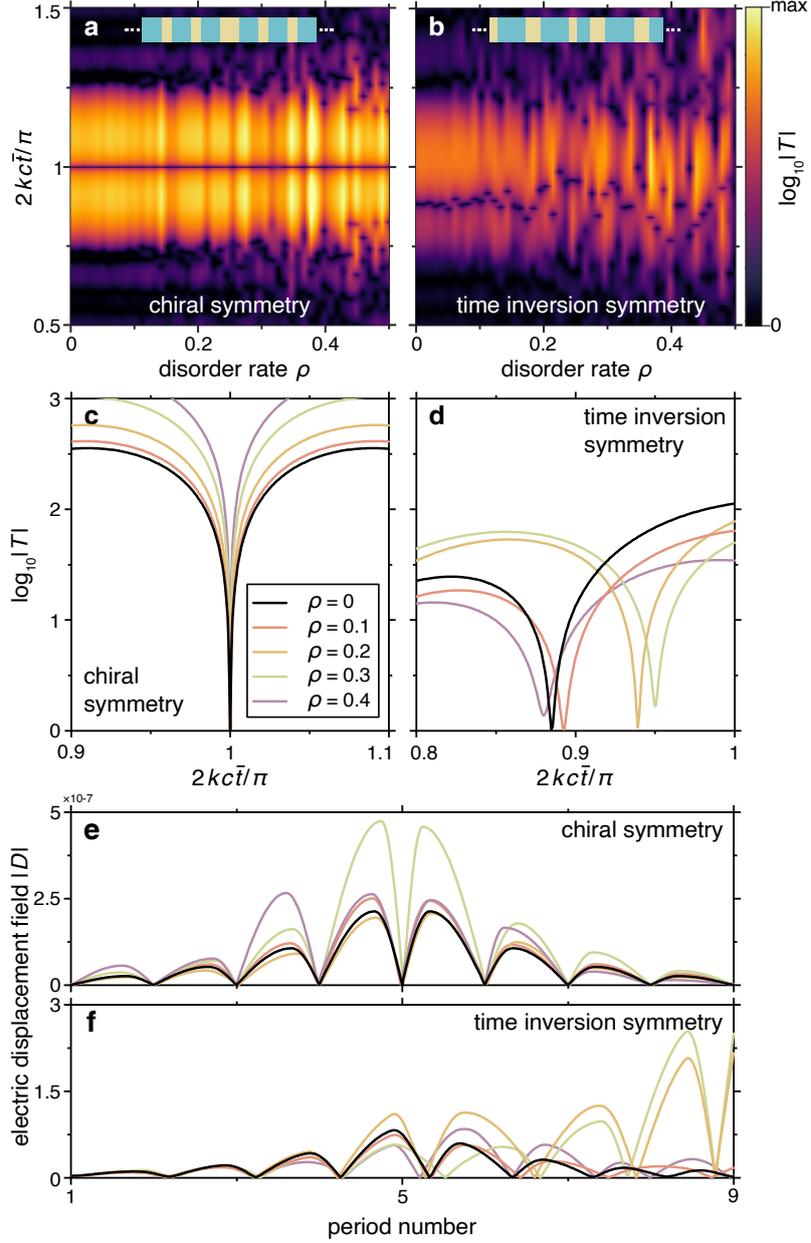

**Figure 4.** Influence of disorders on the temporal topological edge states. (a, b) Transmission spectrum as a function of disorder rate, $\rho$. (c, d) Transmission spectrum for five different disorder rates. The disorder rate is selected as $\rho = 0.1, 0.2, 0.3$ and $0.4$, respectively. In (c) [(d)], the momentum is $2kc\bar{t}/\pi = 1$ [$2kc\bar{t}/\pi = 0.885$], corresponding to the eigenvalue of the temporal topological edge state in the disorder-free system for the two cases. (e, f) Electric displacement distribution as five different disorder rates. In (a, c, e), the TPCs preserve the chiral symmetry whereas in (b, d, f), the TPCs have a time inversion symmetry.

17